\documentclass[%
 reprint,
 amsmath,amssymb,
 aps,
]{revtex4-1}

\usepackage{graphicx}
\usepackage{dcolumn}
\usepackage{bm}
\usepackage[normalem]{ulem}

\begin{document}

\title{Putting Natural Time into Science}
\thanks{To be published as chapter in ''From Astrophysics to Unconventional Computing'', Springer 2019}%

\author{Roger White}
\affiliation{%
 Department of Geography, Memorial University of Newfoundland, St. John's, CANADA \\ roger@mun.ca\\
}%

\author{Wolfgang Banzhaf}

\affiliation{
 Department of Computer Science and Engineering, Michigan State University, East Lansing, MI, USA\\ 
 banzhafw@msu.edu\\
}%


\date{November 19, 2018}

\begin{abstract}
This contribution argues that the notion of time used in the scientific modeling of reality deprives time of its real nature. Difficulties from logic paradoxes to mathematical incompleteness and numerical uncertainty ensue. How can the emergence of novelty in the Universe be explained? How can the creativity of the evolutionary process leading to ever more complex forms of life be captured in our models of reality? These questions are deeply related to our understanding of time. We argue here for a computational framework of modeling that seems to us the only currently known type of modeling available in Science able to capture aspects of the nature of time required to better model and understand real phenomena.
\end{abstract}

\maketitle

\section{INTRODUCTION}
\label{sec:Introduction}

Since its origin more than two millennia ago, epistemological thinking in the West has been driven by a desire to find a way to certify knowledge as certain.  Mathematics and logic seemed to offer a template for certainty, and as a consequence, modern science as it emerged from the work of scholars like Galileo and Newton aimed as much as possible for mathematical formalization.  But by the beginning of the twentieth century the certainty of logic and mathematics was no longer an unquestioned truth: it had become a research question.  From paradoxes in the foundations of logic and mathematics to equations with solutions that are in a sense unknowable, certainty had begun to seem rather uncertain.  By the 1930s, as mathematicians and logicians grappled with the problem of formalisation, they found themselves being forced to look beyond formal systems, and to contemplate the possibility that the solution was of a different nature than they had imagined, that the answers lay in the realm of the living, creative world --- the world of uncertainty.  Although they did not systematically pursue this line of thought, we believe that their intuition was essentially correct, and it serves as our starting point.

The basic reason for the failure of the formalisation programme, we contend, has to do with the inherent nature of logic and mathematics, the very quality that makes these fields so attractive as the source of certainty: their timelessness.  Logic and mathematics, as formal systems, exist outside of time; hence their truths are timeless -- they are eternal and certain.  But because they exclude time,  they are unable to represent one of the most fundamental characteristics of the world: its {\it creativity}, its ability to generate novel structures, processes, and entities.  

Bergson \cite{bergson1911}, by the beginning of the twentieth century, was already deeply bothered by the inability of mathematics and formal science to handle the creativity that is ubiquitous in the living world, and he believed that this was due to the use of ``abstract'' time --- a representation of time --- rather than ``concrete'' or real time.  In spite of his perceptive analysis of the problem, however, he saw no real solution.  In effect mathematics and the hard sciences would be unable to address the phenomenon of life because there was no way they could embody real or natural time \footnote{While time as a parameter has been used in mathematical tools, this amounts to merely a representation of time.} in their formal and theoretical structures.  Real time could only be intuited, and intuition fell outside the realm of hard science.  This view had been anticipated by Goethe \cite{amrine1990,miller2012}, who took metamorphosis as the fundamental feature of nature. Implicitly recognising that formal systems could not encompass metamorphosis, he proposed a scientific methodology based on the development of an intuition of the phenomena by means of immersion in them.  A recent echo of Bergson is found in the work of Latour \cite{latour2012}, who argues that existence involves continuous re-creation in real (natural) time, while the representations of science short-circuit this process of constructive transformation by enabling direct jumps to conclusions. He cleverly refers to this time-eliminating aspect of science as ``double clic''.

The solution to the problem of time, we believe, is to re-introduce real, natural time into formal systems.  Bergson did not believe there was any way of doing this, because science was essentially tied to mathematics or other systems of abstraction which seemed always to eliminate real time. Today, however, we can introduce real or natural time into our formal systems by representing the systems as algorithms and executing them on a computer, which because it operates in natural time, introduces natural time into the algorithm. The answer, in other words, lies in computing.  (In this chapter we will normally use the expression \textit{natural time} rather than real time in order to avoid confusion with the common use of the latter expression to refer to something happening immediately, as in "a real time solution".)

This contribution will develop the argument that various difficulties that arise in logic and formal scientific modelling point to the necessity of introducing natural time itself into our formal treatment of science, mathematics, and logic.  We first discuss some of the difficulties in logic and scientific theory that arise either from a failure to include time in the formal system, or if it is included, from the way it is represented. We then provide  examples of the explanatory power that becomes available with the introduction of natural time into formal systems.  Finally, the last part of the contribution offers an outline of the way forward.

\section{THE ROLE OF TIME IN MATHEMATICS AND SCIENCE}

By the end of the nineteenth century the question of the certainty of mathematics itself was being raised.   This led initially to efforts to show that mathematics could be formalised as a logical system.  Problems in the form of paradox soon emerged in this programme, and those difficulties in turn led to attempts to demonstrate that the programme was at least feasible in principle.  However those attempts also failed when it was proven that mathematics must contain unprovable or undecidable propositions.  The result was that mathematics came to resemble an archipelago of certainties surrounded by a sea of logically necessary uncertainty.  Moreover, with the discovery of the phenomenon of deterministic chaos emerging in the classic three body problem, uncertainties emerged even within the islands of established mathematics.  Meanwhile, in physics, while thermodynamics had long seemed to be somehow problematic because of the nature of entropy, at least it produced deterministic laws.  During the second half of the 20th century, however, it was shown by Prigogine \cite{prigogine1981,prigogine1984} and others \cite{nicolis1977} that these laws are special cases, and that there are no laws governing most phenomena arising in such systems, because the phenomena of interest arise when the systems are far from thermodynamic equilibrium, whereas the laws describe the equilibrium state.  At the same time, in biology, there was a growing realisation that living systems exist in the energetic realm in which the traditional laws of thermodynamics are of only limited use.  And with the discovery that DNA is the genetic material of life, much of biology was transformed into what might be termed the informatics of life.

All of these developments have in common that they introduced, irrevocably and in a radical way, uncertainty and unpredictability into mathematics and science.  They also have in common that they arose from attempts to ensure certainty and predictability by keeping time, real time, out of the formal explanatory systems.  To a large extent, in the practice of everyday science, these developments have been ignored.  Science continues to focus on those areas where certainty seems to be attainable.  However, many of the most interesting and important problems are ones that grow out of the uncertainties that have been uncovered over the past century: problems like the origin and nature of life, the nature of creative processes, and the origin of novel processes and entities. These tend to be treated discursively rather than scientifically. However, we believe that it is now possible to develop formal treatments of problems like these by including time --- real, natural time rather than any formal representation of it --- in the explanatory mechanism.  This will also mean recognising that a certain degree of uncertainty and unpredictability is inherent in any scientific treatment of these problems.  But that would be a strength rather than a failure, because uncertainty and indeterminacy are inherent characteristics of these systems, and so any explanatory mechanism that doesn't generate an appropriate degree of unpredictability is a misrepresentation.  Creativity itself cannot be timeless, but it relies on the timeless laws of physics and chemistry to produce new phenomena that transcend those laws without violating them.  In the next four sections we discuss in more detail the role of time in the treatment of paradox, incompleteness, uncertainty, and emergence in order to justify the necessity of including natural time itself rather than formal time in our explanatory systems. 

\subsection{Paradox}

By the end of the 19th century the desire to show that mathematics is certain had led to Hilbert's programme to show that mathematics could be recast as a syntactical system, one in which all operations were strictly ``mechanical'' and semantics played a minimal role.  In this spirit, Frege, Russell and Whitehead reduced several branches of mathematics to logic, thus apparently justifying the belief that the programme was feasible.  However, Russell's work produced a paradox in set theory that ultimately raised doubts about it. Russell's paradox asks if the set of all sets that do not contain themselves contains itself.  Paradoxically, if it doesn't, then it does; if it does, then it doesn't.  Several solutions have been proposed to rid logic of this paradox, but there is still some debate as to whether any of them is satisfactory \cite{irving2016}.  

The paradox seems to emerge because of the time-free nature of logic.  If we treat the process described in the analysis as an algorithm and execute it, then the output is an endless oscillation. Does the set contain itself?  First it doesn't, then it does, then it doesn't, \dots.   There is no contradiction.   This oscillation depends on our working in discontinuous time --- in this case the time of the computer's clock.  We can in principle make the clock speed arbitrarily fast, but if we could go to the limit of continuous time in operating the computer we would see the paradox re-emerge as something analogous to a quantum superposition: since the superposition endures it doesn't depend on time, and independent of time, the value is always both 'does' and 'doesn't'.  However, if we were to observe the state of the system at a particular instant, the superposition would collapse to a definite but arbitrary value --- ``does'' or ``doesn't'' --- just as Schroedinger's famous cat is only definitively alive or dead when the box is opened.

Of course a computer must work with a finite clock speed and so the paradox cannot appear. This resolution of the paradox by casting it as an algorithm to be executed in natural time emphasizes the difference between the existence of a set and the process of constructing the set, a difference that echoes Prigogine's \cite{prigogine1981}, \cite[p. 320]{prigogine1984} distinction between being and becoming.  In the algorithmic construction of the set, the size of the set oscillates between $n$ and $n+1$, sizes that correspond to ``doesn't'' and ``does''.  Most well known paradoxes are similar to Russell's in that they involve self-reference.  However, Yablo \cite[pp. 24-25]{yanofsky2013} has demonstrated one that is non-self-referential; but it, too, yields an oscillation (true, false, true, false \dots) if executed as an algorithm.

\subsection{Incompleteness}

Whereas Russell discovered a paradox that cast doubt on the possibility of demonstrating that mathematics is a purely syntactical system, both G{\"o}del and Turing came up with proofs that even to the extent that it \textit{is} syntactical, it is impossible to demonstrate that it is. G{\"o}del showed that even for a relatively limited deductive system there were true statements that couldn't be proven true within the system.  In order to prove those statements, the system would have to be enlarged in some way, for example with additional axioms.  But in this enlarged system there would again appear true statements that were not provable within it.  This result highlighted the degree to which mathematics as it existed depended not only on axioms and logical deduction --- that is, syntax --- but also on the products of what was called mathematical intuition --- the source of new concepts and procedures introduced to deal with new problems or problems that were otherwise intractable.  In other words, mathematics was progressively extended by constructions based on these intuitions.  The intuitions came with semantic content, which G{\"o}del's result implicitly suggested could not be eliminated, even in principle.  

But there was another problem.  G{\"o}del, like Hilbert, thought of deduction as a mechanical procedure; thus the idea of certainty was closely linked to the idea of a machine operating in a completely predictable manner.  Of course at the time the idea was metaphorical; the deductions would actually be carried out by a person (the person was invariably referred to as a computer), but for the results to be reliable, the person would have to follow without error a precise sequence of operations that would yield the desired result; in other words, for each deduction the person would have to execute an algorithm.  Thus the algorithm was implicitly part of the logical apparatus used to generate mathematics.  It was clear to Turing and Church that a formal understanding of the nature of these algorithms was necessary.  To that end Turing formulated what is now known as the Turing machine (at the time, of course --- 1937 --- it was purely conceptual), and at about the same time Church developed an equivalent approach, the lambda calculus.  An algorithm was considered legitimate as a procedure if it could be described as a Turing machine or formulated in the lambda calculus.  Each algorithm corresponded to a particular Turing machine. A Turing machine was required to have a finite number of states, and a proof calculated on the machine would need to finish in a finite number of steps.  By proving that it could not in general be shown whether or not algorithms would execute in a finite number of steps (the halting problem), Turing demonstrated, in results analogous to G{\"o}del's, that some propositions were not provable.  Church arrived at the same result using his lambda calculus.  The results of G{\"o}del, Turing, and Church showing that it is impossible to prove that a consistent mathematical system is also complete are complementary to Russell's discovery of the paradox in set theory, which suggests that an attempt to make a formal system complete will introduce an inconsistency.

Turing presumably imposed the restrictions of finite machine states and finite steps in order to ensure that Turing machines would be close analogues of the recursive equations which were the standard at that time for computing proofs. These restrictions on the Turing machine were necessary conditions for proof, but as his results showed, they did not amount to sufficient conditions.  In making these assumptions he effectively restricted the machines to producing time-free results.  This was entirely reasonable given that his goal was to formalize a procedure for producing mathematical results which were themselves timeless.  Nevertheless he found the restrictions to be somewhat problematic, as did Church and G{\"o}del, because they meant that a Turing machine could not capture the generation of the new concepts and procedures that flowed from the exercise of mathematical intuition by mathematicians. There was much informal discussion of this issue, including possibilities for circumventing the limitations.  Turing suggested that no single system of logic could include all methods of proof, and so a number of systems would be required for comprehensive results.  Stating this thought in terms of Turing machines, he suggested a multiple machine theory of mind --- mind because he was still thinking of the computer as a person, and the machine as the algorithm that the person would follow mechanically.  The multiple machine idea was that Turing machines would be chained, so that different algorithms would be executed sequentially, thus overcoming some of the limitations of the simple Turing machine.  How would the sequence be determined?  Initially the idea was that it would be specified by the mathematician, but subsequently other methods were proposed, ranging from stochastic choice to a situation in which each machine would choose the subsequent machine.  Eventually Turing proposed that learning could provide the basis of choice. He observed that mathematics advances by means of mathematicians exercising mathematical intuition, which they then use to create new mathematics, a process he thought of as learning.  He then imagined a machine that could learn by experience, by means of altering its own algorithms. (Copeland and Shagrir, 2015)  

Time had been kept out of mathematics by defining it as consisting only of the achieved corpus of results and ignoring the process by which those results were generated.  Turing and Church took a first step toward including the process by formalizing the treatment of the algorithms by which proofs were derived.  But this did not seem to them (or others) to be sufficient, because it did not capture the deeply creative nature of mathematics as seen in the continual introduction of new concepts and procedures that established whole new areas of mathematics.  We might interpret the journey from Turing algorithm to learning as an implicit recognition of the necessity of natural time in mathematics.  On the other hand, while Turing had formalized the proof process in terms of a machine which would have to operate in natural time, the machine was defined in such a way that the results that it produced would be time free.  Thus in postulating strategies like chained Turing machines to represent learning, he probably assumed that the results would also be time free. However, if mathematics is understood to include the process by which it is created, it will have to involve natural time, even if the result of that creative process is time free. While Turing spoke of learning, a more appropriate term might be creativity, and creativity --- the emergence of something new --- necessarily involves time. 

\subsection{Uncertainty}

But is mathematics, in the narrower sense of established results, really entirely timeless?  Perhaps. But there is at least one small part of it --- deterministic chaos --- that seems to be trying to break free and take up residence in natural time.  The phenomenon was discovered by Henri Poincar{\'e} at the beginning of the twentieth century as he attempted to solve the Newtonian three-body problem.  An equation that exhibits deterministic chaotic dynamics can, up to a point, be treated as a timeless structure and its properties investigated using standard mathematical techniques (some invented by Poincar{\'e} for this purpose). It has been shown, for example, that the attractor is a fractal object, and therefore infinitely complicated.  As a consequence the solution trajectory cannot be written explicitly.  It can, however, be calculated numerically --- but only up to a point: since the attractor, as a fractal, is infinitely complex, we can never know exactly where the system is on it, and hence cannot predict the future states of the system.

Take one of the simplest possible cases, the difference equation version of the logistic function:
\begin{equation}
\begin{split}
X_{t+1} = r X_t (1 - X_t) \\ \mbox{ with } 0<X_t<1 \mbox{ and }0<r<4
\end{split}
\end{equation}
Solving the equation for $X$ as a function of $t$ we have
\begin{equation}
X^{*} = 1-1/r
\end{equation}

This solution is stable for $r \le 3$; otherwise it is oscillatory. For $r > 3.57$ approximately, the oscillations are infinitely complex; i.e. the dynamics are chaotic. Since the solution cannot be written explicitly, to see what it looks like we calculate successive values of $X$, while recognizing that these values become increasingly approximate, and soon become entirely arbitrary --- that is, they become unpredictable even though the equation generating them is deterministic.  This can be dismissed as simply due to the rounding errors that result from the finite precision of the computer, but it is actually a consequence of the interaction of the rounding errors with the fractal nature of the attractor.  The rate at which the values evolve from precise to arbitrary is described by the Lyapunov exponent.  So while the system may look well defined and timeless from an analytical point of view because the attractor determines the complete behaviour of the system, in fact the attractor is unknowable analytically, and can only be known (and only in a very limited way) by iterating the equation, or by other iterative techniques such as the one developed by Poincar{\'e}.  The iterations take place in time, and so it seems that at least some of our knowledge of the behaviour of the equation necessarily involves natural time.  Note that a physical process characterised by chaotic dynamics must also be unpredictable in the long run, because as the process ``executes'' it will be following a fractal attractor, and the physical system that is executing the process, being finite, will be subject to ``rounding errors'', which in effect act as a stochastic perturbation.  In other words, the resolution that can be achieved by the physical system is less than that of the attractor.  This is the case with the three body problem that Poincar{\'e} was working on, and the reason that the planetary trajectories of the solar system are unpredictable at timescales beyond a hundred million years or so.  It is also, in Prigogine's (1997) view, the fundamental reason for the unpredictability of thermodynamic systems at the microscopic level (Laplace's demon cannot do the math well enough), and also a necessary factor in macroscopic self-organization in that the unpredictability permits symmetry breaking---Prigogine calls this order by fluctuations. 

With chaotic systems, then, we lose the promise that mathematics has traditionally provided of a precise, God's eye view over all time, and thus of certainty and predictability.  We are left only with calculations in natural time that give us rapidly decreasing accuracy and hence increasing unpredictability.  But from some points of view this is not necessarily a problem.  As Turing speculated in his discussion of learning, learning requires trial and error, which would be pointless in a perfectly predictable world, and specifically it requires an element of stochasticity.  Many others have made the same observation -- that stochasticity seems to be a necessary element in any creative process.  In physical systems stochastic perturbations are the basis of the symmetry breaking by which systems become increasingly complex and organized.  Chaotic dynamics may thus play a positive role by providing a necessary source of stochasticity in physical, biological, and human systems. 

Physics, with the major exception of thermodynamics (together with two very specific cases in particle physics: CPT and a case of a heavy-fermion superconductor \cite{schemm2014}), is characterized by laws that are ``time reversible'' in the sense that they remain valid if time runs backward.  In other words, time can be treated as a variable, t, and the laws remain valid when $-t$ is substituted for $t$.  This is referred to by some (e.g. \cite{smolin2013}) as spatialized time, because we can travel in both directions in it, as we can in space.  Spatialized time is a conceptualization and representation of natural time, whereas natural time is the time in which we and the world exist, independently of any representation of it. Spatializing time is thus a way of eliminating natural time by substituting a model for the real thing. The physics of spatialized time is essentially a timeless physics, since we have access to the entire corpus of physical laws in the same sense that we have access to the entire body of timeless mathematics. 

The fact that time can be treated as a variable permits the spectacularly accurate predictions that flow from physical theory:  the equations can be solved to show the state of the system as a function of time, and thus the state at any particular time, whether past, present or future.  This physics is in a deep sense deterministic.  This is true even of quantum physics, where, as Prigogine \cite{prigogine1997} points out, the wave function evolves deterministically; uncertainty appears only when the wave function collapses as a result of observation.  The determinism of spatialized time is the basis of Einstein's famous remark that ``for us convinced physicists, the distinction between past, present and future is an illusion, although a persistent one'' (as quoted in \cite[p. 165]{prigogine1997}). His point was that time as we experience it flowing inexorably and irreversibly is an illusion; in relativistic space-time, the reality that underlies our daily illusory existence, we have access to all times.   However, Prigogine \cite{prigogine1997} points out that the space-time of relativity is not necessarily spatialized; that is just the conventional interpretation.  In any case, because it is apparently timeless, the physics of quantum theory and relativity is understood to represent our closest approximation to certain knowledge of the world.

Thermodynamics represents a rude exception to this timelessly serene picture.  Here time has a direction, and when it is reversed the physics doesn't work quite the same way.  In the forward direction of time, the entropy of an isolated system increases until it reaches the maximum possible value given local constraints.  In this sense the system is predictable.  But when time is reversed, so that entropy is progressively lowered, the system becomes unpredictable, because, as Prigogine showed, when the entropy of a system is lowered, an increasing number of possible states appears, states that are macroscopically quite distinct but have similar entropy levels. But only one of these can actually exist, and in general we have no certain way of knowing which one that will be.  The same phenomenon appears in reversed computation.  In other words the reversed time future is characterized by a bifurcation tree of possibilities. Its future is open; it is no longer deterministic or fully predictable, but rather path dependent.  This discussion of reversed time futures applies to isolated systems, the ones for which thermodynamic theory was developed.  However, we do not live in an isolated system.  Our planet is bathed in solar energy, which keeps it far from equilibrium, and we supplement this with increasing amounts of energy from other sources.  Thus our open-system world is equivalent to a reversed time, isolated-system world.  It is a world of path dependency and open futures of a self-organizing system.

The open futures of these systems is a source of unpredictability or uncertainty, just as is the uncertainty arising from chaotic dynamics, and the two work together.  In both of these situations in which unpredictability appears, time continues to be treated as a variable, but in the case of far from equilibrium systems the behaviour is time asymmetric: if we treat decreasing entropy as equivalent to time reversal, physical systems are deterministic in $+t$ but undetermined in $-t$. In the case of chaotic systems, while the \textit{process} may be mathematically deterministic in both $+t$ and $-t$, the \textit{outcome} is undetermined for both directions of time.  In both cases, since a mathematical treatment of the phenomenon is of limited use, the preferred approach is computational.  This is not just a pragmatic choice.  It reflects the poverty of spatialized time compared to the possibilities offered by real time.  Whereas physics, with the major exception of thermodynamics, is based on the assumption that spatialized time captures all the characteristics of time that are essential for a scientific understanding of the world, natural time involves no assumptions.  It is simply itself.  A computer can only implement an algorithm step by step, in natural time.  As a consequence, algorithms as they are executed do not depend on any conceptualization or representation of time beyond a working assumption that time is discontinuous or quantized, rather than continuous, an assumption imposed by the computer's clock.

\subsection{Emergence}

Far from equilibrium, self-organizing systems are the ones that we live in; our planet is essentially a spherical bundle of such systems.  The dynamics of plate tectonics is driven by energy generated by radioactive decay in the earth's core; the complex behaviour of the oceans and atmosphere is driven by the flux of solar energy; and life itself, including human societies, also depends on the continuous input of energy from the sun. Self-organization is a kind of emergence --- it is the process by which an organized structure or pattern emerges from the collective dynamics of the individual objects making up the system, whether these are molecules of nitrogen and oxygen organizing themselves into a cyclonic storm or individual people moving together to form an urban settlement. However, as the phrase self-organization suggests, there is no prior specification of the form that is to emerge, and because of the inherent indeterminacy of far-from equilibrium systems, there is always a degree of uncertainty as to exactly what form will appear, as well as where and when it will emerge. These forms are essentially just patterns in the collection of their constituent particles or objects.  Unlike their constituent objects, they have no existence as independent entities, and they cannot, simply as patterns, act on their environment --- in other words, they have no agency.  For this reason the emergence of self-organized systems is called soft or weak emergence. 

Strong emergence, on the other hand, refers to the appearance of new objects, or new types of objects, in the system.  We can identify three levels of strong emergence:

\begin{enumerate}
\item
In high energy physics, forces and particles emerge through symmetry breaking.  Unlike the increasing energy input required to drive self-organization, this process occurs as free energy in the system decreases and entropy increases.

\item
At relatively moderate energy levels, physical systems produce an increasing variety of chemical compounds.  These molecules have an independent existence and distinctive properties, like a characteristic colour of solubility in water, that are not simply the sum of the characteristics of their constituent atoms.  They also have a kind of passive agency: for example, they can interact with each other chemically to produce new molecules with new properties, like a new colour.  Of course they can also interact physically, by means of collisions, to produce weak emergence, for example in the form of a convection cell or a cyclonic storm.  But chemical reactions can result in the simultaneous occurrence of both strong and weak emergence, as when reacting molecules and their products generate the macroscopic self-organized spiral patterns of the Belosov-Zhabotinsky reaction.  The production of a particular molecule may either use or produce free energy, i.e. it may be either entropy increasing or entropy decreasing.  

\item
Also at relatively moderate energy levels, living systems emerge through chemical processes, but also through self-assembly of larger structures (cells, organs, organisms).  The key characteristic of this level of strong emergence is that the process is initiated and guided by an endogenous model of the system and its relationship with its environment.  While in (1) and (2) emergence is determined by the laws of physics, in this case it is determined by the relevant models working together with the laws of physics and chemistry.  We include in living systems the meta-systems of life such as ecological, social, political, technological, and economic systems.
\end{enumerate}

It is this third kind of strong emergence, the kind that depends on and is guided by models, that is the focus of our interest.  Nevertheless the weak emergence of self-organizing systems remains important in the context of strong emergence, because a process of strong emergence, as in the case of the development of a fertilised egg into a mature multi-cellular individual, often makes use of local self-organization.  Furthermore, self-organized structures are often the precursors of individuals with agency, making the transition by means of a process of reification, as when a self-organized settlement is incorporated as a city, a process that endows it with independent agency.  In general, while self-organized systems are forced to a state of lower entropy by an exogenously determined flux of energy, living systems create and maintain their organized structures in order to proactively import energy and thus maintain a state of lower entropy.  The causal circularity is a characteristic of such systems.

It is this third kind of strong emergence, the kind that depends on and is guided by models, that is the focus of our interest.  Nevertheless the weak emergence of self-organizing systems remains important in the context of strong emergence, because a process of strong emergence, as in the case of the development of a fertilised egg into a mature multi-cellular individual, often makes use of local self-organization.  Furthermore, self-organized structures are often the precursors of individuals with agency, making the transition by means of a process of reification, as when a self-organized settlement is incorporated as a city, a process that endows it with independent agency.  In general, while self-organized systems are forced to a state of lower entropy by an exogenously determined flux of energy, living systems create and maintain their organized structures in order to proactively import energy and thus maintain a state of lower entropy.  The causal circularity is a characteristic of such systems.

Model based systems are a qualitatively new type.  The models provide context dependent rules of behaviour that supplement the effects of the laws of physics and chemistry.  Of course we can always reduce the structures that act as the models to their basic chemical components in order to understand, for example, the chemical structure of DNA or the chemistry of the synapses in a network of neurons, and there are good reasons for doing this: it allows us to understand the underlying physical mechanisms by which the model --- and by extension the system of which it is a part --- functions.  But this reduction to chemistry and physics tells us nothing about how or why the system as a whole exists.  These questions can only be answered at the level of the model considered as a model, because it is the model that guides the creation and functioning of the system of which it is a part.  In other words, the reductionist programme reveals the syntax of the system, but tells us nothing of the semantics.  It is the rules of behaviour of the system as a whole, rules provided by the model, that determine the actions of the system in its environment, and thus, ultimately, its success in terms of reproduction or survival.  Part of the semantic content of the model is therefore the teleonomic goal of survival.  The teleonomy is the result of the evolutionary process that produced the system.  In this sense evolution is the ultimate source of semantics: as Dobzhansky said in the famous title of his paper, "Nothing in biology makes sense except in the light of evolution" \cite{dobzhansky1973}.  The mathematical biologist Robert Rosen \cite{rosen1991}\cite{rosen2000} speculated that life, rather than being a special case of physics and chemistry, in fact represents a generalization of those fields, in the sense that a scientific explanation of life would reveal new physics and chemistry.  In other words the models inherent in living systems could be seen as new physics and chemistry: they introduce semantics as an emergent property of physico-chemical systems. 

\subsubsection{Models}

An interesting and useful definition of life, due to Rosen \cite{rosen1991}, is that life consists of entities that contain models of themselves, that is, entities that exist and function by virtue of the models they contain.  The most basic model is that coded in DNA. But neural systems also contain models, some of them, as we know, very elaborate. And of course some models are stored in external media such as books and computers.  These three loci of models correspond to the three worlds of Karl Popper:  World 1 is the world of physical existence, World 2 corresponds to mental phenomena or ideas; and World 3 consists of the externally stored and manipulated representations of the ideas.  Worlds 2 and 3 are not generally considered by scientists to be constituents of the world that science seeks to explain.  However, as Popper points out, they are in fact part of it, and exert causal powers on World 1 \cite{popper1982}.  The implication is that a scientific understanding of biological and social phenomena requires not just an analysis at the level of physical and chemical causation, but also consideration of the causal role of meaning, or more specifically, meaning as embodied in models.  Thus semantics re-enters the picture in a fundamental way.

A model that is a part of a living system must be a formal structure with semantics, not just syntax.  It can function as a model only by virtue of its semantic content, since in order to be a model it must represent another system, a system of which, in the case of living organisms, it is itself usually a part.  The modelled system thus provides the model with its semantic content.  As Rosen points out, this contradicts the orthodox position of reductionist science (and in particular of Newtonian particle physics) that ``every material behaviour [can] be...reduced to purely syntactical sequences of configurations in an underlying system of particles'' \cite[p.68; see also p.46ff]{rosen2000}. 

Non-living systems lack semantics; they might thus be characterised as identity models of themselves, or \textit{zero-order models}.  Models associated with living organisms (e.g. DNA or an idea of self) would then be \textit{first order models}. And some scientific models, those that are models of models (e.g. a mathematical model of DNA), would be \textit{second order models}. This chapter is concerned with first order models.   \\

We propose the following definition: \\
  
\textbf{\textit{a}} is a first order model of \textbf{\textit{A}}, i.e. \textbf{\textit{a}} is a \textit{functional representation} of \textbf{\textit{A}} (\textbf{{a \sout{r} A}}), if

\begin{enumerate}
\item  \textbf{\textit{a}} is a structure (not just a collection) of  appropriate elements in some medium, whether chemical (e.g. a DNA molecule composed of amino acids), cellular (e.g. a synaptic structure in a network of neurons), or symbolic (e.g. a program composed of legitimate statements in some programming language).
\item The structure \textbf{\textit{a}} can act as an algorithm when executed on some suitable machine \textbf{\textit{M}}, where \textbf{\textit{M}} may be either separate from \textbf{\textit{A}} (e.g. a computer running a model of an economic system), or some part of \textbf{\textit{A}} (e.g. a bacterial cell running the behavioural model coded in its DNA);
\item The output of the algorithm corresponds to or consists of some characteristics of \textbf{\textit{A}}.  Specifically:

		(a) Given a suitable environment, \textbf{\textit{a}} running on\textbf{\textit{ M}} can \textit{create} a new instance of \textbf{\textit{A}} (e.g. in the environment provided by a warm egg, the DNA being run by the egg cell containing the DNA can create a new instance of the kind of organism that produced the egg).
        
		(b)  \textbf{\textit{a}} can \textit{guide the behaviour} of \textbf{\textit{A}} in response to certain changes in the state of the environment (e.g. on the arrival of night, go to your nest; if inflation is greater than 3 percent, raise the interest rate).
        \item 3(a) and 3(b) are \textit{evolved} (or in human World 3 systems, \textit{designed}) capabilities or functions that in general serve to maximise the chance of survival of \textbf{\textit{A}}.
        \item If \textbf{\textit{A}} is a living organism, \textbf{{ \sout{r}}} is an emergent property of the underlying physical and chemical systems.
\end{enumerate}

First (and higher) order models are essentially predictive.  Although the output of \textbf{\textit{a}} when executed on \textbf{\textit{M}} is literally a response to a current condition \textbf{\textit{c}} which represents input to \textbf{\textit{a}}, because of the evolutionary history of \textbf{\textit{a}} that brought it into existence, the behaviour of \textbf{\textit{A}} in response to \textbf{\textit{a}} is actually a response to a future, potentially detrimental, condition \textbf{\textit{c}}' \textit{predicted} by \textbf{\textit{a}}; in other words, on the basis of the current condition \textbf{\textit{c}}, \textbf{\textit{a}} predicts that \textbf{\textit{c}}' will occur, and as a consequence produces a response in\textbf{\textit{ A}} intended to prevent the occurrence of \textbf{\textit{c}}' or mitigate its impact.  Thus \textbf{\textit{a}} acts as a predictive algorithm, and guides the behaviour of \textbf{\textit{A}} on the basis of its predictions.  

The model \textbf{\textit{a}} is thus rich in time.  It involves both past time (the time in which it evolved) and future time (the time of its prediction), as well as, during execution, the natural time of the present. This reminds us of Bergson's \cite[p.20]{bergson1911} observation regarding natural ("concrete") time:  "the whole of the past goes into the making of the living being's present moment." Only if we know already about evolution as a process can we see the three times present in \textbf{\textit{a}}. Only by virtue of being such a system ourselves do we have the ability to perceive its purpose.

In this three-time aspect, \textbf{\textit{a}} as it represents \textbf{\textit{A}} is fundamentally different from a purely chemical or physical phenomenon in the conventional sense. It has semantic content which would be eliminated by any possible reduction to the purely mechanical causation of chemical and physical events.   From a reductionist standpoint we would see only chemical reactions, nothing of representation, purpose, past, or anticipation.  On the other hand, since \textbf{\textit{a}} does actually have this semantic content, that content must emerge in the chemical system itself.  It does so by virtue of the relationship between the part of the system that constitutes \textbf{\textit{a}} and the larger system that is being modelled, just as a molecular property like solubility in water emerges from interactions among the atoms making up the molecule.  In this sense Rosen was correct that life represents a radical extension of chemistry and physics: at no point do we require a vital principle or a soul to breathe semantics, or even life, into chemistry.

In the specific case of DNA, as a molecule it is essentially fixed from the point of view of the organism: over that timescale, as a molecule, it is timeless.  But natural time appears as the organism develops following conception, when various genes are turned on or off, and this behaviour continues in the fully developed organism as its interactions with the environment are guided by various contingently activated combinations of genes.  In that sense DNA acts as a model that changes as a result of its interactions with the modelled system of which it is a part.  This is reminiscent of the chained Turing machines proposed by Turing to permit creativity.  Neural models, in contrast, lack a comprehensive fixed structure analogous to that of DNA; they are open ended and develop or change continually as a result of interactions with their host organism and the environment.  But in both cases, as a computational system, life is essentially a case of open ended computation.

\subsubsection{Information}

The model of a system represents information, and its role in the functioning of the system depends on its being treated as information by the system.  Note that this is information in the sense of semantics, or meaningful information, rather than Shannon information, which is semantics-free and represents information capacity or potential information.  In other words, we could say that while semantics represents the content or meaning of information, Shannon information represents its quantity, and syntax represents its structure.  Shannon information is maximised when a system is in its maximum entropy state.  In the case of a self-organized system, the macro-scale pattern constrains the behaviour of the constituent particles so that the system's entropy and hence its Shannon information is less than it would be if its particles were unconstrained by the self-organized structures. 

We do not know of a measure of semantic information; it seems unlikely that such a measure could even be defined.  Nevertheless, it seems that the model is the means by which semantic content emerges from syntax.  We speculate that it is ultimately the teleonomic nature of living systems that populates the vacant lands of Shannon information with the semantics of meaning-laden information. A model embedded in a living system does not simply represent some aspect of another system; it does so purposefully.  In living systems, the function of the model is to guide the behaviour of the system of which it is a part, and it does this by predicting future states of both the system and its environment.  System behaviour thus depends to some degree on the anticipated future state of the system and its environment --- i.e. the behaviour is goal directed.  In contrast, in the case of traditional feedback systems, behaviour depends on the current state of the system and its environment.  We note the apparent irony that life, a system that depends for its origin and evolution on uncertainty, nevertheless depends for its survival on an ability to predict future states.  In fact, it requires a balance of predictability and unpredictability. In Langton's \cite{Langton1990} terms, it exists on the boundary between order and chaos.  

\subsubsection{Agency}

First order models emerged with life in an evolutionary process, one in which the model both depends on and facilitates the persistence of the system of which it is a part.  The model thus necessarily has a teleonomic quality --- its purpose is ultimately to enhance the likelihood of its own survival and that of the host system that implements it.  To this end, the model endows its host system with agency --- i.e. it transforms the system into an agent that can act independently.  The relationship between model and evolutionary process, the basis of strong emergence, seems fundamental: each seems to be necessary for the other. This is in a sense the basic assumption of the theory of biological evolution.  In contrast, a self-organized system, the result of weak emergence, does not act independently to ensure its own persistence.  Living systems, by virtue of their agency, act to maintain themselves in a state of low entropy.


\subsection{Creative Algorithms}

The models that guide the generation and behaviour of living systems are necessarily self-referential, since they are models of a system of which they are an essential part.  This means that they cannot be represented purely as mathematical structures.  However, if the mathematical structures  are appropriately embedded in algorithms being executed in natural time, the problem disappears.  Nevertheless, the definition of algorithm remains crucial.  Rosen, with deep roots in mathematics, was never quite able to resolve the problems arising from self-reference because he worked with Turing's definition of algorithm; this is clear when he claims, repeatedly, that life is not algorithmic.  But as we have noted, the Turing machine was defined in such a way as to produce only results that are consistent with time-free mathematics.  To generate that mathematics, the Turing machine must be supplemented by a source of learning or creativity.  Learning and creativity are essential characteristics of living systems, as is the appearance of new entities with agency, which learning and creativity make possible. Consequently, a formal understanding of life must include a formal treatment of learning, creativity and strong emergence.  That  requires algorithms that transcend Turing's definition. It requires algorithms that are able to model their own behaviour and alter themselves on the basis of their models of themselves.  Using a computer operating in natural time to execute only Turing algorithms is like insisting on using three dimensional space to do only two dimensional geometry: it is a colossal waste of capacity as well as a refusal to consider the unfolding world of possibilities that emerge in natural time. 

\section{FURTHER EXPLORATIONS ON THE ROLE OF TIME IN SCIENCE}

We have gotten so used to the concept of creativity and completely new solutions to problems, or to inventions that make our life easier and are introduced the first time, that we tend to overlook the principle aspect of creating new things. 

In the daily processes of synthetic chemistry, for example, new molecules are generated every day by combining existing molecules into new combinations. Given the enormous extent of the combinatorial space of chemistry, we have to presume that some of those are created the very first time. 

If some of these compounds are stable and created today in the Universe for the first time -- note we speak of actual realization of material compounds, as opposed to the mere possibility of their existence being ``discovered'' -- they come with a time stamp of today. Thus, every material substance or object has in some way attached to it a time stamp of when it or its earlier copies first appeared in the Universe. Time, therefore, is of absolute importance to everything that exists and is able to characterize it in some way. 

Can we make use of that in the Sciences? Here, we want to look at the two sciences that provide modeling tools for others to use in their effort to model the material universe, mathematics and computer science.

\subsection{Mathematics}

We have already mentioned that mathematics uses the concept of time (if at all) in a spatial sense. This means, time can be considered as part of a space that can be traversed in all directions. Notably, it can be traversed backward in time! But mathematics is actually mostly concerned about the unchanging features of the objects and transformations it has conceptualized. Thus, it glosses over, or even ignores changes in features, as they could prevent truth from getting established. For instance, a mathematical proof is a set of transformations of a statement into the values ``true'' or ``false'', values that are unchanging and not dynamic. This reliability is its strength. Once a statement is established to be true, it is accepted into the canon of mathematically proven statements, and can serve as an intermediate for other proof transformations. 

But what about a mathematics of time? How would such a mathematics look like? We don't know yet, perhaps because the notion of time is something many mathematicians look at with suspicion, and rather than asking how such a mathematics would look like, they ask themselves whether time exists at all and how they can prove that it does not exist --- except as an illusion in our conciousness \cite{barbour2001}. Although Science has always worked like that --- ignoring what it cannot explain and focusing on phenomena it can model and explain --- we have reached a point now where we simply cannot ignore the nature of time any more as a concept that is key  to our modeling of natural systems. 

So let's offer another speculation here. We said before that every object in the universe carries a property with it we can characterize as a time stamp, stating when it first appeared. This is one of its unalienable properties, whether we want to consider it or not. So how about imagining that every mathematical object and all the statements and transformations in mathematics would come with the feature of a time stamp? In other words, besides its other properties, an object, statement or transformation would carry a new property, the time when it was first created. This would help sort out some of the problems when trying to include the creation of mathematics into mathematics itself. It would actually give us a way of characterizing how mathematics is created by mathematicians. The rule would be that new objects, statements and transformations can only make use of what is already in existence at the time of their own creation.

Once we have achieved such a description, can we make a model of the process? Perhaps one of the natural things to ask is whether it would be possible to at least guess which objects, statements or transformations could be created next?
The situation is a reminder of the ``adjacent possible'' of Kaufmann who proposed that ecological systems inhabit a state space that is constantly expanding through accessing ``adjacent'' states that increase its dimensionality.
What this includes is a notion that 
only what interacts with the existing (which we can call ``the adjacent'') could be realized next. Everything else would be a {\it creatio ex nihilo} and likely never be realized.

Here is an example: Suppose we have a set of differential rate equations that describe a system at the current state. For simplicity, let's assume that all the variables of the system carry a time stamp of this moment. Suppose now that we want to introduce a new variable, for another quantity that develops according to a new differential rate equation. Would it make sense to do that even without any coupling of this new variable to the existing system? We don't think it would. In fact, the very nature of our wish to introduce this variable has to do with its interaction with the system as it is currently described. Thus, introducing a variable that can describe the adjacent possible has at least to have {\it some} interaction with the current system. 

Dynamic set theory \cite{liu1993} is an example of how this could work. Dynamic set theory was inspired by the need to deal with sets of changing elements in a software simulation. Mathematically, normal sets are static, in that membership in a set does not change over time. But dynamic sets allow just that: Sets can be defined over time intervals $T$, and might contain certain elements at certain times only. For example, if you have a set of elements 
\begin{equation*}
X = \{a_1, a_2, a_3, b_1, b_2, b_3, c_1, c_2, c_3 \}
\end{equation*}
we can assign specific dynamic sets to a time interval $T$
as follows
\begin{equation*}
A^T = \{ (t_1, \{ a_1, b_1, c_1 \}), (t_2, \{a_2, b_1 \}),(t_T, \emptyset)  \}
\end{equation*}
and
\begin{equation*}
B^T = \{ (t_1, \{ a_1, a_2 \}), (t_3, \{a_3, c_1, c_3 \}),(t_T, \emptyset)  \}
\end{equation*}
We can then manipulate these sets using set operations, for instance:
\begin{equation}
A^T \cap B^T = \{ (t_1, \{ a_1 \}),(t_T, \emptyset)  \}
\end{equation}
or
\begin{equation}
\begin{split}
A^T \cup B^T = \{ (t_1, \{ a_1, a_2, b_1, c_1 \}),  (t_2, \{a_2, b_1 \}), \\ (t_3, \{a_3, c_1, c_3 \}), 
(t_T, \emptyset)  \}
\end{split}
\end{equation}
We can see here that each of these elements is tagged with a particular time at which they are part of the dynamic set, and can take part in set operations for that particular moment.\footnote{Note that we have skirted the issue of how to measure time, and how to precisely determine a particular moment and its synchronous counterparts in other regions of the Universe. For now, we'd stick to classical time and assume a naive ability to measure it precisely.} A generalization of set theory is possible to this case. Our hope is that --- ultimately --- mathematics will be able to access the constructive, intuitional aspects of its own creation. Once we have assigned the additional property of time/age to mathematical objects, perhaps its generative process can be modeled. 

Another example of mathematical attempts at capturing time in mathematics is real-time process algebra \cite{wang2002}. The idea of this approach is to try to describe formally what a computational system is able to do, in particular its dynamic behavior. This project of formalization was generalized under the heading of ``denotational mathematics'' \cite{wang2014,wang2015}.

These are all interesting attempts to capture the effect of time within the framework of Mathematics, but they fall short of the goal, because they are descriptive in nature, i.e. they are not generative and able to create novel structures, processes or variables themselves.  

\subsection{Computer Science}

Computers allow the execution of mathematical models operationalized as algorithms. But as we have seen from the discussion in this chapter, mathematics currently deals with spatialized time, not real, natural time. Thus, if we were to only aim at simulating mathematical models, we do not need natural time. This is indeed Turing's definition of an algorithm, restricted exactly in the way required to make sure that it cannot do anything that requires natural time, so that the computer executing a Turing algorithm is only doing what, in principle, timeless mathematics can do.  
Here, instead, we aim for algorithms to execute on machines that need to go beyond traditional mathematical models. 

We need to provide operations within our algorithms that allow for modification of models. Let us briefly consider how variables (potential observables of the behavior of a [simulated] model) are realized in a computer: They are handled using the address of their memory location. Thus if we allow memory address manipulations in our algorithms, like allocating memory for new variables, or garbage collection (for variables/memory locations that have fallen out of use), we should be able to modify at least certain aspects of a model (the variables). Since the address space of a computer is limited, memory locations can be described by integer numbers in a certain range, so we are able to modify them during execution.

Of course, variables are but one class of entities that need to be modified from within the code. Reflective computer languages allow precisely this type of manipulation \cite{smith1982}. Reflection describes the ability of a computer language to modify its own structure and behavior. Mostly interpreted languages have been used for reflection, yet more modern approaches like {\it SELF} offer compiling capabilities, based on an object-oriented model. As Sobel and Friedman write: ``Intuitively, reflective computational systems allow computations to observe and modify properties of their own behavior, especially properties that are typically observed only from some external, meta-level viewpoint'' \cite{sobel1996}. What seems to make {\it SELF} particularly suitable is its ability to manipulate methods and variables in the same framework. In fact, there is no difference in {\it SELF} between them. Object classes are not based on an abstract collection of properties and their inheritance in instantiation, but on prototype objects, object copy and variation. We believe that {\it SELF} allows an easier implementation of an evolutionary system than other object-oriented languages. 

Susan Stepney's work \cite{stepney2011} in the context of the CoSMoS project provides a good discussion of the potential of reflective languages to allow to capture emergent phenomena through self-modification.  In order for a self-modifying system not to sink into a chaotic mess, though, we probably shall need again to time stamp the generation of objects.

However, the open-ended power of those systems might only come into its own when one of the key aspects of natural time is respected as well --- the fact that one cannot exit natural time. This calls for systems that are not terminated. Natural open-ended processes like scientific inquiry or economic activity or biological evolution {\it do not allow} termination and restart. While objects in those systems might have a limited lifetime, entire systems are not ``rebooted''. Instead, new objects have to be created and integrated into the dynamics of the existing systems. 

We return here to a theme already mentioned with Turing machines: The traditional idea of an algorithm, while having to make use of natural time during its execution as a step-by-step process, attempts to ignore time by requiring the algorithm to halt. Traditional algorithms are thus {\it constructed} to halt for their answer to be considered definitive. This, in fact, makes them closed system approaches to computation, as opposed to streaming processes, that analyze data continuously and provide transient answers at any time \cite{dodig2011}. We might want to ask: What are the requirements for systems that do not end, i.e. do not exit natural time? \cite{banzhaf2016}

\subsection{OTHER SCIENCES}


In this contribution we do not have enough space to discuss in detail how natural phenomena as encountered in simple and complex systems can inform the corresponding sciences --- which attempt to model those phenomena (physics, chemistry, biology, ecology and economy) --- about natural time. But we believe it is important to emphasize that a clear distinction should be made between our modeling attempts and the actual phenomena underlying them. In the past, there were times when the model and the reality were conceptually not separated. For instance, the universe was considered like clockwork, or later as a steam engine, and even later as a giant computer. All of these attempts to understand the universe mistook the underlying system for its metaphor.  \\

\section{CONCLUSION}

Our argument here is not that the Universe is a giant computer \cite{fredkin2003}, preferably running an irreducible computation \cite{wolfram2002}. This would interchange the actual system with the model of it. Rather, our argument is that time is so fundamental to the Universe that we need tools (computers) and formalisms (algorithms) that rely on natural time to be able to faithfully model its phenomena. We believe that there are many phenomena in the natural and artificially made world making use of novelty, innovation, emergence, or creativity, which have resisted modeling attempts with current techniques. We think those phenomena are worth the effort to change our concepts in order to accommodate them into our world view and allow us to develop models. As hard as it might be to do that, what would Science be without taking stock of what is out there in the world and attempting to incorporate it in our modelling efforts?

\vspace{3mm}
This essay was written on the occasion of the Festschrift for Susan Stepney's 60th birthday. It is dedicated to Susan, whose work has been so inspiring and deep.


\bibliography{NaturalTime}

\end{document}